\documentclass[10pt,preprint] {aastex}
\usepackage{amssymb,amsmath,gensymb}
\usepackage{graphicx,epsfig,natbib,color}
\usepackage{lscape}
\usepackage{graphicx}
\usepackage{epsfig}
\usepackage{natbib}
\usepackage{rotating}

\slugcomment{Accepted for publication in ApJ}

\begin{document}
\title{Investigating the Conditions of the Formation of a Type II Radio Burst on 2014 January 8}

\author{W. Su,\altaffilmark{1,2,3}, X. Cheng,\altaffilmark{1,3}, M. D. Ding,\altaffilmark{1,3}, P. F. Chen,\altaffilmark{1,3}, Z. J. Ning,\altaffilmark{2}, H. S. Ji, \altaffilmark{2}}

\affil{$^1$ School of Astronomy and Space Science, Nanjing University, Nanjing 210023, China}\email{xincheng@nju.edu.cn; dmd@nju.edu.cn}
\affil{$^2$ Key Laboratory of Dark Matter and Space Astronomy, Purple Mountain Observatory, Nanjing 210008, China}
\affil{$^3$ Key Laboratory for Modern Astronomy and Astrophysics (Nanjing University), Ministry of Education, Nanjing 210023, China}

\begin{abstract}
It is believed that type II radio bursts are generated by shock waves. In order to understand the generation conditions of type II radio bursts, in this paper, we analyze the physical parameters of a shock front. The type II radio burst we selected was observed by Siberian Solar Radio Telescope (SSRT) and Learmonth radio station and was associated with a limb CME occurring on 2014 January 8 observed by the Atmospheric Imaging Assembly (AIA) on board the Solar Dynamics Observatory (SDO). The evolution of the CME in the inner corona presents a double-layered structure that propagates outward. We fit the outer layer of the structure with a partial circle and divide it into 7 directions from -45$^\circ$ to 45$^\circ$ with an angular separation of 15$^\circ$. We measure the outer layer speed along the 7 directions, and find that the speed in the direction of -15$^\circ$ with respect to the central direction is the fastest. We use the differential emission measure (DEM) method to calculate the physical parameters at the outer layer at the moment when the type II radio burst was initiated, including the temperature ($T$), emission measure ($EM$), temperature ratio ($T_{d}/T_{u}$), compression ratio ($X$), and Alfv\'{e}n Mach number ($M_{A}$). We compare the quantities $X$ and $M_{A}$ to that obtained from band-splitting in the radio spectrum, and find that this type II radio burst is generated at a small region of the  outer layer that is located at the sector in 45$^\circ$ direction. The results suggest that the generation of type II radio bursts (shock) requires larger values of $X$ and $M_{A}$ rather than simply a higher speed of the disturbance.
\end{abstract}


\section{Introduction}

MHD shocks are an important phenomenon in the solar atmosphere, which can accelerate and heat charged particles effectively \citep{Schwartz1988,Mann1995}. They can be generated by coronal mass ejections \citep[CMEs,][]{Vrsnak2008}, reconnection jets \citep{Magara2000}, fast expanding loops \citep{Su2015}, or the ``CME-streamer" interaction in the lower corona \citep{Eselevich2015}. A coronal shock from the Sun can produce a significant impact on the space weather and human beings. Type II radio bursts represent a typical feature of coronal shocks in radio wavelengths \citep{Wild1950,Zheleznyakov1970}. Generally, type II radio bursts are believed to originate from the shock front, but it is difficult to determine the exact position along the shock front. \citet{Kouloumvakos2014} calculated the compression ratio of the shock, and found that the type II radio burst can originate from the whole sheath region between the CME leading edge (LE) and the shock front. However, most authors considered that the source region of the type II radio burst is at a small region of the shock front \citep{Bain2012,Zimovets2012,Carley2013,Zucca2014b,Zimovets2015}. Previous works suggested that the type II radio burst is most likely generated at the nose of the shock fronts \citep{Bemporad2011}, because the shock speed at the nose is usually higher than that at the flank. However, \citet{Reiner2003} found that type II radio bursts can be generated in the high density coronal streamer when the shock flank travels through it. Similar type II radio bursts with such a generation mechanism have been reported successively \citep{Cho2007a,Feng2012,Kong2015}.

The key question on where the source region of the type II radio burst in the shock front is, or what physical conditions are needed for the generation of the type II radio, is still open. There are some studies on this question. As is well known, Alfv\'{e}n Mach number is a key parameter to describe the strength of a shock, and the type II radio bursts can be generated more easily when the Alfv\'{e}n Mach number is larger \citep{Shen2007}. \citet{Gopalswamy2009} analyzed the data from Solar Terrestrial Relations Observatory \citep[{\it STEREO};][]{Kaiser2008} and found that the type II radio bursts are generated at the height where the Alfv\'{e}n speed is the minimum. \citet{Cho2013} found a type II burst that originates in the low corona (0.08 R$_{\odot}$) when the shock passes through high density coronal loops. They attributed it to the low Alfv\'{e}n speed because of the high density. \citet{Kozarev2011} used the Differential Emission Measure (DEM) method to get the compression ratio ($X$) and the temperature ratio of the shock. It was found that the temperature does not change obviously before and after the shock passes through the region, and a lower limit of $X$ was obtained for the shock. \citet{Zucca2014a} made a density map using extreme ultraviolet (EUV) data from the Atmospheric Imaging Assembly \citep[AIA;][]{Lemen2012} on board the Solar Dynamics Observatory \citep[{\it SDO};][]{Pesnell2012} and polarized brightness data from Large Angle and Spectrometric Coronagraph \citep[{\it LASCO};][]{Brueckner1995} on board the Solar and Heliospheric Observatory ({\it SOHO}). Combining with the potential-field source-surface (PFSS) model, they also calculated the Alfv\'{e}n speed map. \citet{Bemporad2014} obtained the vector magnetic field at the pre- and post-shock regions by measuring the profile and speed of the shock and assuming a radial direction of the magnetic field in the field of view of LASCO/C2. \citet{Susino2015} derived 2-dimensional (2D) maps of the electron density, the compression ratio, and the Alfv\'{e}n Mach number in the region where the shock swept based on the LASCO/C2 observation.

The purpose of this work is to study the physical properties of the coronal shock responsible for a type II burst based on combined observations including multi-wavelength EUV data and the radio dynamic spectrum. Besides, we try to get the properties of the coronal shocks from the radio dynamic spectrum. More than ten years ago, \citet{Vrsnak2002} proposed a method to get the compression ratio of a shock from the band-splitting of type II radio bursts. \citet{Ma2011} calculated the compression ratio and the temperature of the 2010 June 13 event from band-splitting. However, the obtained values of these parameters are larger than those yielded by the DEM method. The reason may be the over-simplified assumption that the emission measure (EM) per unit length along the line of sight (LOS) is uniform, which actually is not, especially in the situation where different structures exist along the LOS. To obtain a more consistent result, it is thus necessary to combine two different methods, like the DEM analysis of the multi-wavelength data and the band-splitting in radio dynamic spectrum, as mentioned above, to investigate the properties of the coronal shock. To this end, we select a typical event of type II radio burst on 2014 January 8 and study its generation conditions (the properties of the coronal shock).

This paper is organized as follows. Observations are presented in Section \ref{sect:Observations}, and data analysis and results are described in Section \ref{sect:Data Analysis}. We then have some discussions in Section \ref{sect:Discussion}, followed by the conclusion in Section \ref{sect:Conclusions}.

\section{Observations}
\label{sect:Observations}

In order to study the properties around the shock from the start time of the type II radio burst, we select an event with relatively more data from different instruments and a full coverage in frequency and time. First, such an event should have band-splitting in the radio spectrum, so that we can calculate the compression ratio ($X$); Second, the eruption of the event must be at the limb so that we can reliably apply the DEM method to get the EM and temperature; Third, the eruption must be in the field of view of AIA when the type II radio burst starts, so that we can compare the results from the radio spectrum and those from the DEM method.

According to the three selection criteria above, we choose a type II radio burst that occurred on 2014 January 8 and was observed by Learmonth Observatory in Australia. A few minutes before the type II radio burst, an M3.6-class flare occured on the west limb of the Sun, in the active region NOAA 11947. The flare started at 03:39 UT, and peaked at 03:46 UT in GOES soft X-ray flux. At 04:12 UT, a CME appeared in the field-of-view (FOV) of LASCO/C2 coronagraph. The angular width of the CME is 108$^\circ$. The time sequence indicates that the shock responsible for the type II radio burst is most likely a CME-driven one.

The start frequency of the type II radio burst is higher than 300 MHz in the first harmonic frequency, out of the range of the Learmonth radio station. Therefore, we need to seek radio data with higher frequencies from other radio observatories data to show the burst in higher frequencies. Fortunately, this burst was also observed by ALAMTY, GAURI, MRT1, OOTY, and SSRT radio stations. We use the SSRT data in higher frequencies because the burst is the clearest in its spectrum. Here, we only use the radio spectrum in 45--441 MHz, although SSRT has frequencies above GHz ( the data of SSRT are from \url{http://www.e-callisto.org/}). By comparison, the frequency range of the data by Learmonth is 25--180 MHz, and the time resolution is 3~s. As shown in Figure \ref{4-spec-LEAR-SSRT}, the radio dynamic spectrum, with the combination of Learmonth and SSRT data, covers the whole range of the type II radio burst.

The frequency-drifting pattern in Figure \ref{4-spec-LEAR-SSRT} is a typical type II radio burst. There are obvious branches of fundamental and first harmonic frequencies in the dynamic spectrum. In the first harmonic frequency branch, the type II radio burst started at about 03:45:00 UT with a frequency of about 350~MHz, and ended at 03:57:00 UT with a frequency of 70~MHz. In particular, from 03:46:20 UT, the band-splitting can be seen clearly in the first harmonic frequency. The type II radio burst drifts from higher frequencies to lower frequencies smoothly, indicating a gradual drop of the coronal plasma density on the path of the shock. At the beginning of the type II radio burst, there also appears a type III radio burst, indicating the presence of energetic electrons.

On the other hand, as shown in Figure 2, there is a pileup region ahead of an eruption flux rope that propagates outward as seen in the extreme ultraviolet (EUV) images observed by {\it SDO}/AIA during the type II radio burst. The pileup region is most obvious in 193 {\AA} and 211 {\AA} (middle and right columns in panel (a) of Figure 2), less obvious in 171 {\AA} and 335 {\AA}, and marginally visible in 94 {\AA} and 131 {\AA} (left column in panel (a) Figure 2) wavelengths. Note that the 94 {\AA} and 131 {\AA}~emissions indicate high temperature plasma. One can also see from panel (b) of Figure 2 that the pileup region is double-layered in the 193 {\AA} and 211 {\AA} passbands. The two layers, an inner one and an outer one, can be clearly distinguished. The green arrows point to the inner layer (IL), while the pink arrows point to the outer layer (OL). The stand-off distance between them is narrow in the direction of eruption, but it is much wider at the two flanks. The distance between the IL and the OL becomes larger with propagation, implying that the OL moves faster than the IL. These characteristics have been observed in other CME-driven shocks \citep{Ontiveros2009, Chen2011, Gopalswamy2012, Eselevich2012, Kouloumvakos2014, Lee2014}, suggesting that the OL may be the front of a CME-driven shock. The double layers appear at 03:45 UT in 193 {\AA} and 211 {\AA}, and then propagate outward. At 03:46 UT (the beginning of the type II radio burst), the IL and OL are both arc-shaped. The OL begins to deviate from a circular shape at 03:47 UT, when a small hump (indicated by the white arrow in Figure 2) appears at a sector of the OL. At 03:48 UT, the OL moves out of the fields of view of {\it SDO}/AIA.

\section{Results}
\label{sect:Data Analysis}

In order to identify whether the OL of the structure is a CME-driven shock front or not, we further analyze the properties of different regions around the OL. To this end, we divide the OL into a number of segments along different directions. \citet{Hundhausen1994} showed that the projected shape of a CME is like an ice cream, the CME speed can be divided into a radial speed and an expansion speed, the latter of which is isotropic in each direction ideally. If the speed of the driver is isotropic, the CME-driven shock should also be isotropic in each direction. Fortunately, for this event, the OL is close to a standard circular arc and visible most obviously in 193 {\AA} and 211 {\AA}. Thus, we fit the OL with a circular arc based on the 193 {\AA} and 211 {\AA} images. We then set the line connecting the center of the solar disk and the center of the fitted circle as the base line (0$^\circ$ direction), and select 7 directions from -45$^\circ$ to 45$^\circ$ with an angular separation of 15$^\circ$ between adjacent directions.

We select 3 moments corresponding to the different stages of the event, from the beginning of the type II radio burst appearing in the dynamic spectrum where the lower and upper frequency branches in the harmonic band are both clear, to a time when the OL propagates out of the FOV in the EUV images by {\it SDO}/AIA, as shown in Figure \ref{4-spec-LEAR-SSRT}. The three selected moments in the dynamic spectrum are 03:46:23 UT, 03:46:48 UT, 03:47:12 UT, respectively. The closest times in the {\it SDO}/AIA 193 {\AA} images are 03:46:18 UT, 03:46:42 UT, and 03:47:06 UT, respectively, with a difference of only several seconds. The fitted shapes of the OL at the three moments in the AIA 193 {\AA} and 211 {\AA} images are shown in Figure~\ref{4-Alldirection}. The centers of the fitted circles for each time, marked as white plus signs in the figure, are (974$\arcsec$, 208$\arcsec$), (975$\arcsec$, 207$\arcsec$), and (1004$\arcsec$, 215$\arcsec$) in the heliocentric coordinates. The centers of the fitted circles are close to each other.

\subsection{Propagation Speeds}

As the centers of the fitted circles at different moments are close to each other, we can consider that the expansion speed is dominant during the propagation of the OL. Likewise, since the OL resembles a circular arc, the expansion speed of the OL is nearly isotropic along each direction. Thus, we use the straight lines as shown in Figure~\ref{4-Alldirection} to trace the OL propagation. We get the time-slice plots for each direction as shown in Figure~\ref{4-V-211}, where any propagating front shows up as a bright ridge. Along the directions close to the base line, the width of the bright ridge increases with time, indicating that the stand-off distance between the OL and IL increases with time. Here we consider the frontal side of the bright ridge as the OL, and the rear side of the ridge as the IL.

Based on the slopes of the bright ridges, we measure the IL and OL speeds from the time-slice images shown in Figure~\ref{4-V-211}. The OL speeds are 594 km s$^{-1}$, 637 km s$^{-1}$, 636 km s$^{-1}$, and 577 km s$^{-1}$ in the 0$^{\circ}$, 15$^{\circ}$, 30$^{\circ}$, and 45$^{\circ}$ directions, respectively. The OL speed does not vary too much in these 4 directions. However, the OL speed is 798 km s$^{-1}$ in the direction of -15$^{\circ}$, faster than those in the other directions. The speeds of the \textbf{IL} are 426 km s$^{-1}$, 622 km s$^{-1}$, 463 km s$^{-1}$, 612 km s$^{-1}$, 583 km s$^{-1}$, and 435 km s$^{-1}$, in the directions of -30$^{\circ}$, -15$^{\circ}$, 0$^{\circ}$, 15$^{\circ}$, 30$^{\circ}$, and 45$^{\circ}$, respectively, somewhat lower than the speeds of OL in each direction. Note that we can find only a narrow faint ridge in the -30$^{\circ}$ direction, which may correspond to the IL with a speed of 426 km s$^{-1}$. In the -45$^{\circ}$ direction, we cannot deduce the speed because both the IL and OL are too faint in the running-difference images as shown in Figure~\ref{4-Alldirection}.

\subsection{Parameters Derived from Radio Spectrum}

The fundamental frequency of type II radio bursts is the same as the local plasma oscillation frequency at the shock front. There often appears a band-splitting in the dynamic spectrum of type II radio bursts, in which two similar branches, an upper one and a lower one, drift toward lower frequencies with time. There are several interpretations for this phenomenon. One interpretation suggests that the band-splitting could result from the different parts of the shock front simultaneously when the shock encounters different background plasma with different densities \citep{McLean1967,Holman1983}. Recently, \citet{Eselevich2016} proposed that the band-splitting is caused by the simultaneous propagation of a piston shock and a blast wave. In this model, the two branches are interpreted to be due to different speeds and different propagation directions of the two shocks. In contrast to these models, the most popular interpretation suggests that the upper and lower frequencies of the band-splitting are simultaneously generated by the downstream and upstream of a coronal shock, respectively, because there is a density jump across the shock front \citep{Smerd1974,Smerd1975}. There are still some disputes on the exact mechanisms responsible for the band-splitting \citep[e.g.,][]{Treumann1992,Du2015}. The interpretation proposed by \citet{Smerd1975} is preferred in many recent papers \citep{Cho2007b,Ma2011,Gopalswamy2012,Zimovets2012}.

\begin{figure}
  \center
   \includegraphics[width=12cm,trim = 0 120 0 0]{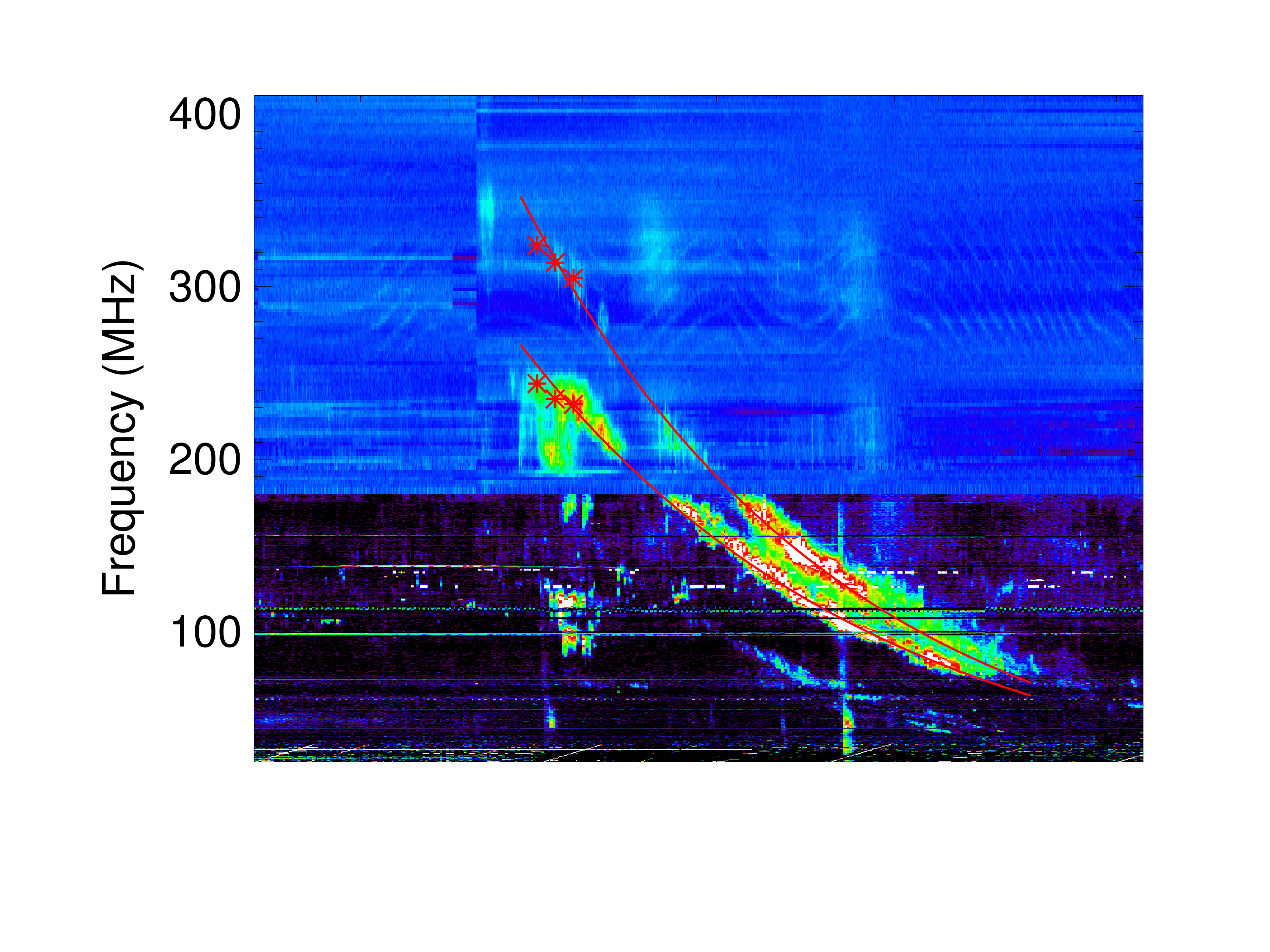}
   \includegraphics[width=12cm,trim = 0 0 0 50]{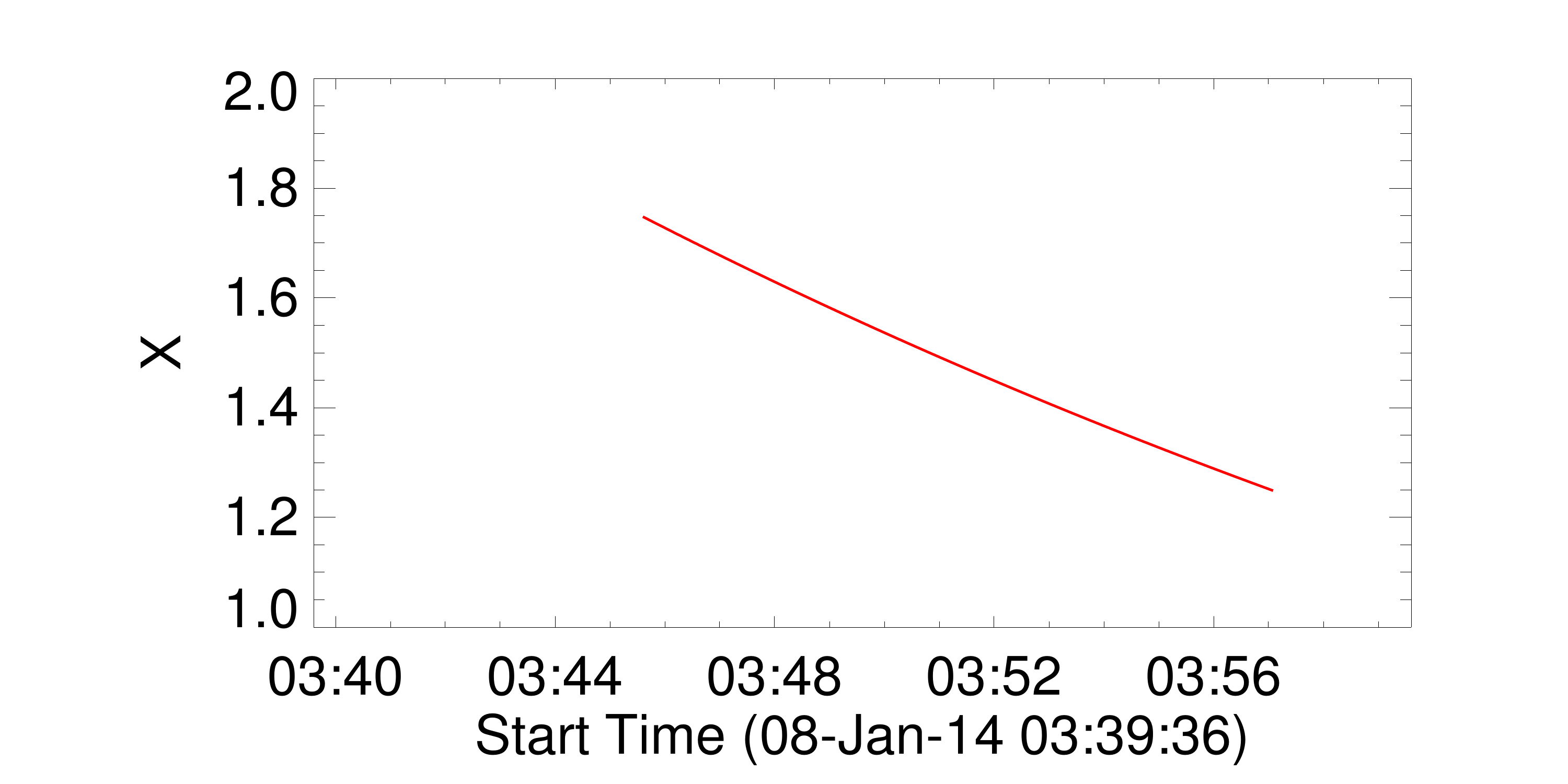}
   \caption{Top panel is dynamic spectrum of the type II radio burst on 2014, January 8. Red lines in the top panel are the fitting curves of the upper and lower branches for the harmonic frequency of the type II radio burst. Bottom panel is the evolution of $X$.}
   \label{4-spec-LEAR-SSRT}
\end{figure}

\citet{Vrsnak2002} proposed a method to obtain the compression ratio of the shock from the band-splitting of a type II radio burst. As the frequency of the type II radio burst ($f$) is related to the plasma number density ($n$) by $f \varpropto \sqrt{n}$, we can get the compression ratio ($X$) of the shock from the upper and lower frequencies in the band-splitting:
\begin{equation}
\label{equation1}
    X = \frac{n_{2}}{n_{1}} = (\frac{f_{U}}{f_{L}})^2,
\end{equation}
where $n_{2}$ and $n_{1}$ are the number densities of the plasma in the downstream and upstream of the shock, respectively, $f_{U}$ is the upper branch frequency and $f_{L}$ is the lower branch frequency of the type II radio burst. From the fitting curves (shown in the top panel of Figure~\ref{4-spec-LEAR-SSRT}) of the upper and lower branches of the band-splitting, we get the evolution of $X$ (shown in the bottom panel of Figure~\ref{4-spec-LEAR-SSRT}) during the entire type II radio burst. It is found that the value of $X$ decreases with time. This is probably because the Alfv\'{e}n wave speed of the background corona increase with height so that the Alfv\'{e}n Mach number decreases as demonstrated by \citet{Gopalswamy2012}.

For this event, we get the upper and lower frequencies of the band-splitting at the three moments, as marked with the asterisks in the dynamic spectrum shown in Figure~\ref{4-spec-LEAR-SSRT}. With Equation (\ref{equation1}), the values of $X$ are found to be 1.77 $\pm$ 0.21 , 1.78 $\pm$ 0.16 , 1.72 $\pm$ 0.18 for $t$ = 03:46:23 UT, 03:46:48 UT, and 03:47:12 UT, respectively. Since the shock is still in the low corona at these moments, the magnetic field can be considered as closed field here; therefore, the shock is likely a perpendicular one \citep{Ma2011} for this event. We can then calculate the Alfv\'{e}n Mach $M_{A}$ number for a perpendicular shock in ideal MHD ($\beta \rightarrow$ 0):
\begin{equation}
\label{equation2}
    M_{A} = \sqrt{\frac{X(X + 5)}{2(4 - X)}}, \\
\end{equation}
For the three moments, the values of $M_{A}$ are derived to be 1.64 $\pm$ 0.18, 1.65 $\pm$ 0.14, and 1.60 $\pm$ 0.15, respectively.

\subsection{Parameters Derived from the DEM method}

\begin{figure}[htbp]

\begin{minipage}[b]{\textwidth}
\centering
\begin{minipage}[b]{0.48\textwidth}
\includegraphics[width = 8 cm,clip = true]{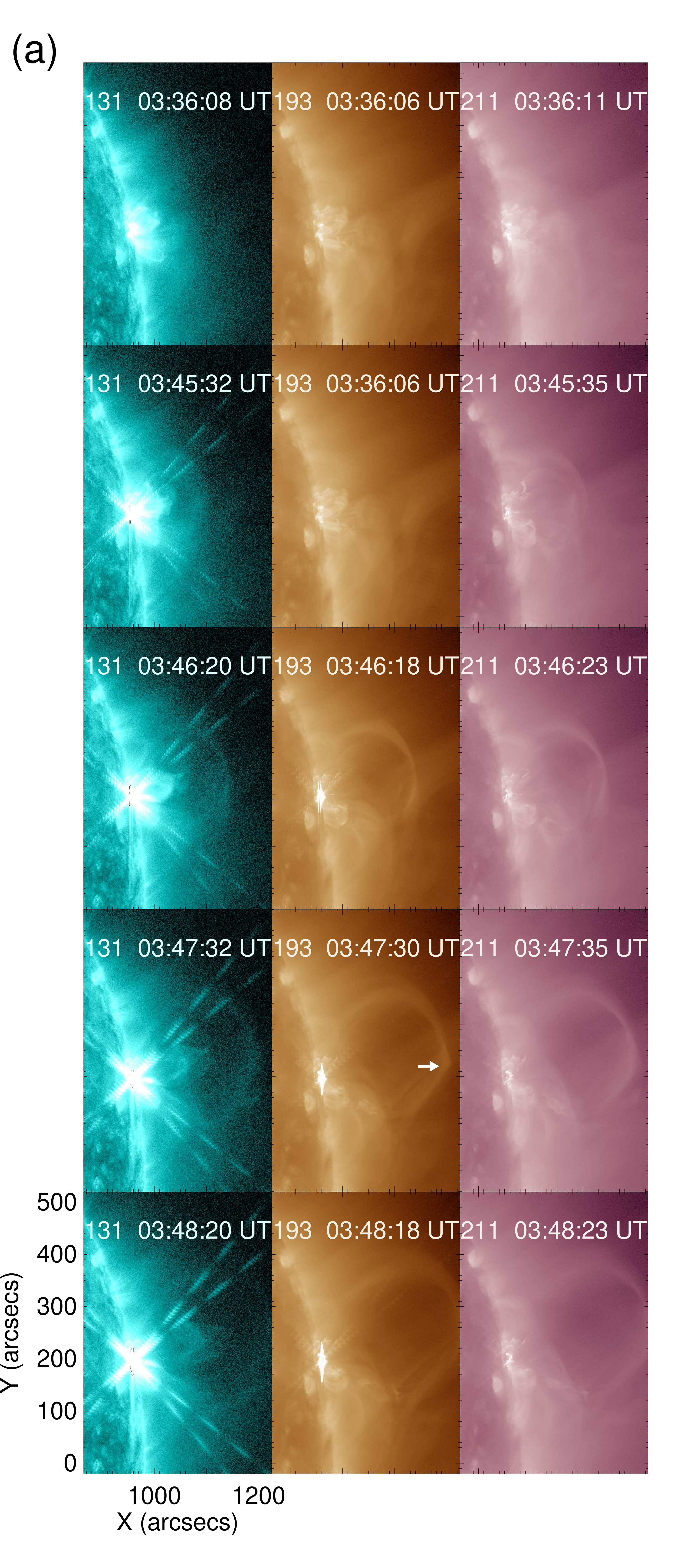}
\end{minipage}
\begin{minipage}[b]{0.48\textwidth}
\includegraphics[width = 9 cm,trim = 0 0 0 0,clip = true]{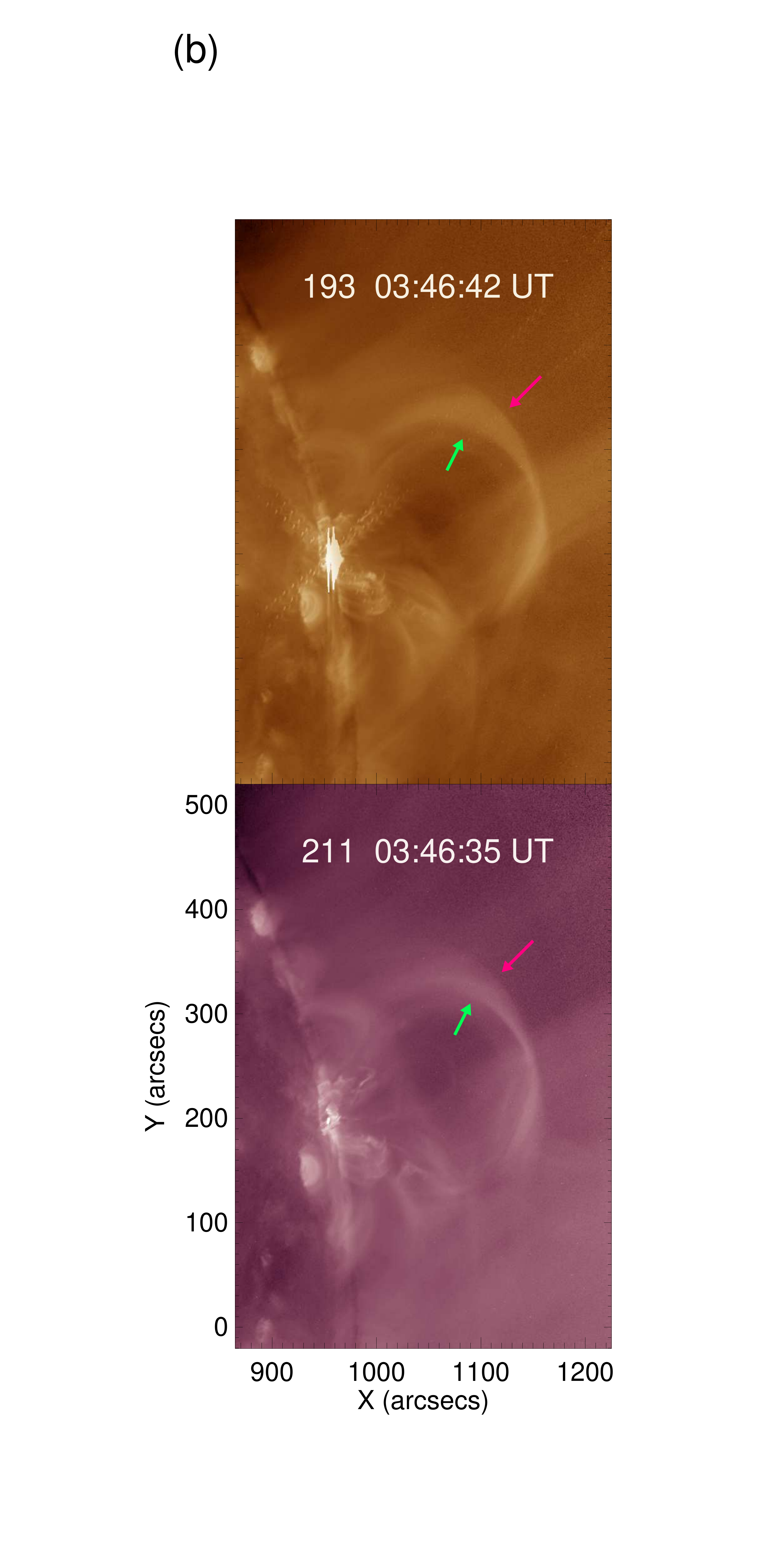}
\end{minipage}
\end{minipage}
\label{4-aia-overview}
\caption{\small{{\it SDO}/AIA images at 131 {\AA} (left), 193 {\AA} (middle) and 211 {\AA} (right) showing the evolution of the event during the type II radio burst. A double-layered structure is seen to propagate outward in 193 {\AA} and 211 {\AA} images. The inner layer is shown as green arrows, while the outer layer is shown as pink arrows.}}
\end{figure}

\begin{figure}

\begin{minipage}[b]{\textwidth}
\centering
\begin{minipage}[b]{0.25\textwidth}
\includegraphics[width = 4.1cm,trim = 0 120 0 80,clip = true]{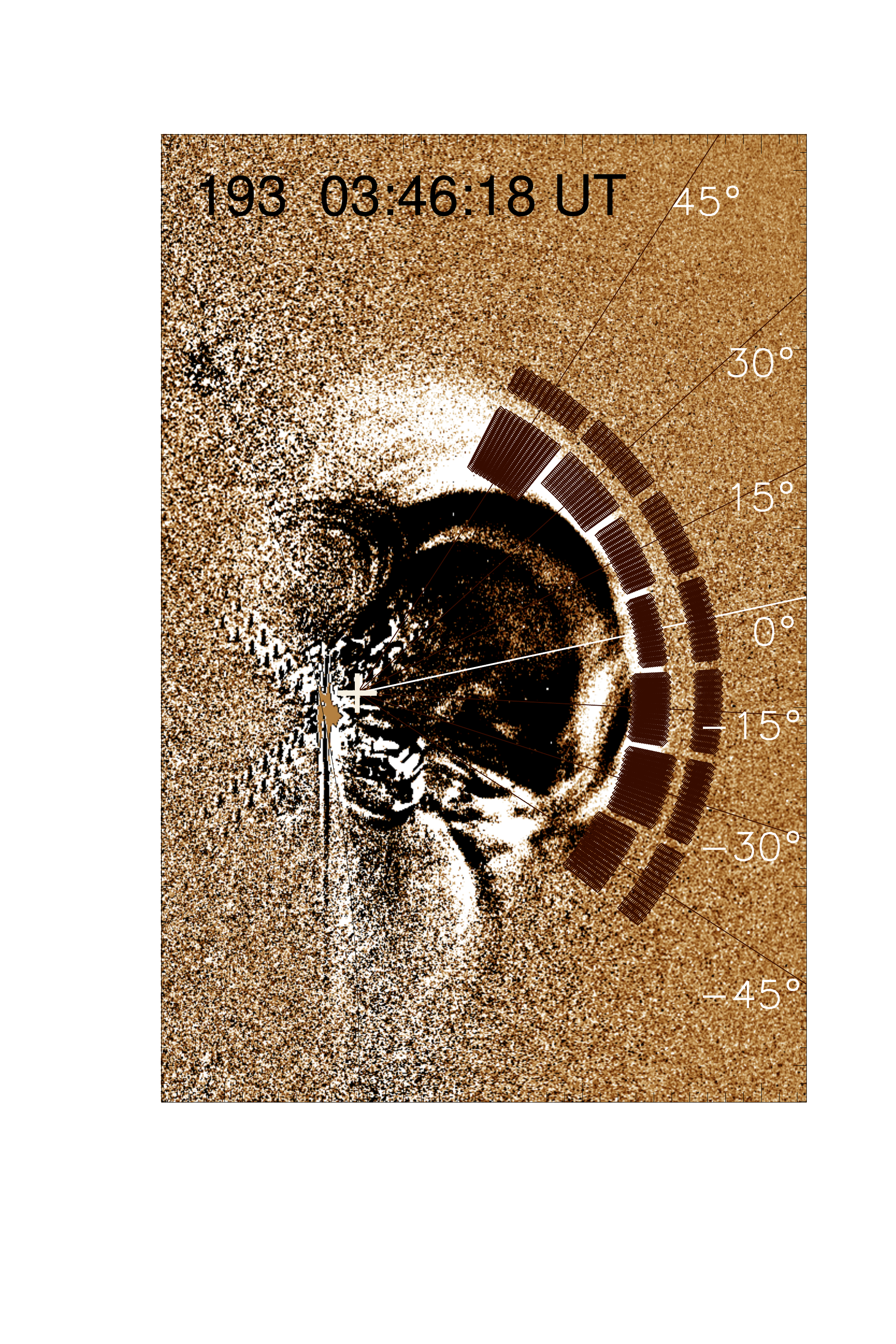}
\end{minipage}
\begin{minipage}[b]{0.25\textwidth}
\includegraphics[width = 4.1cm,trim = 0 120 0 80,clip = true]{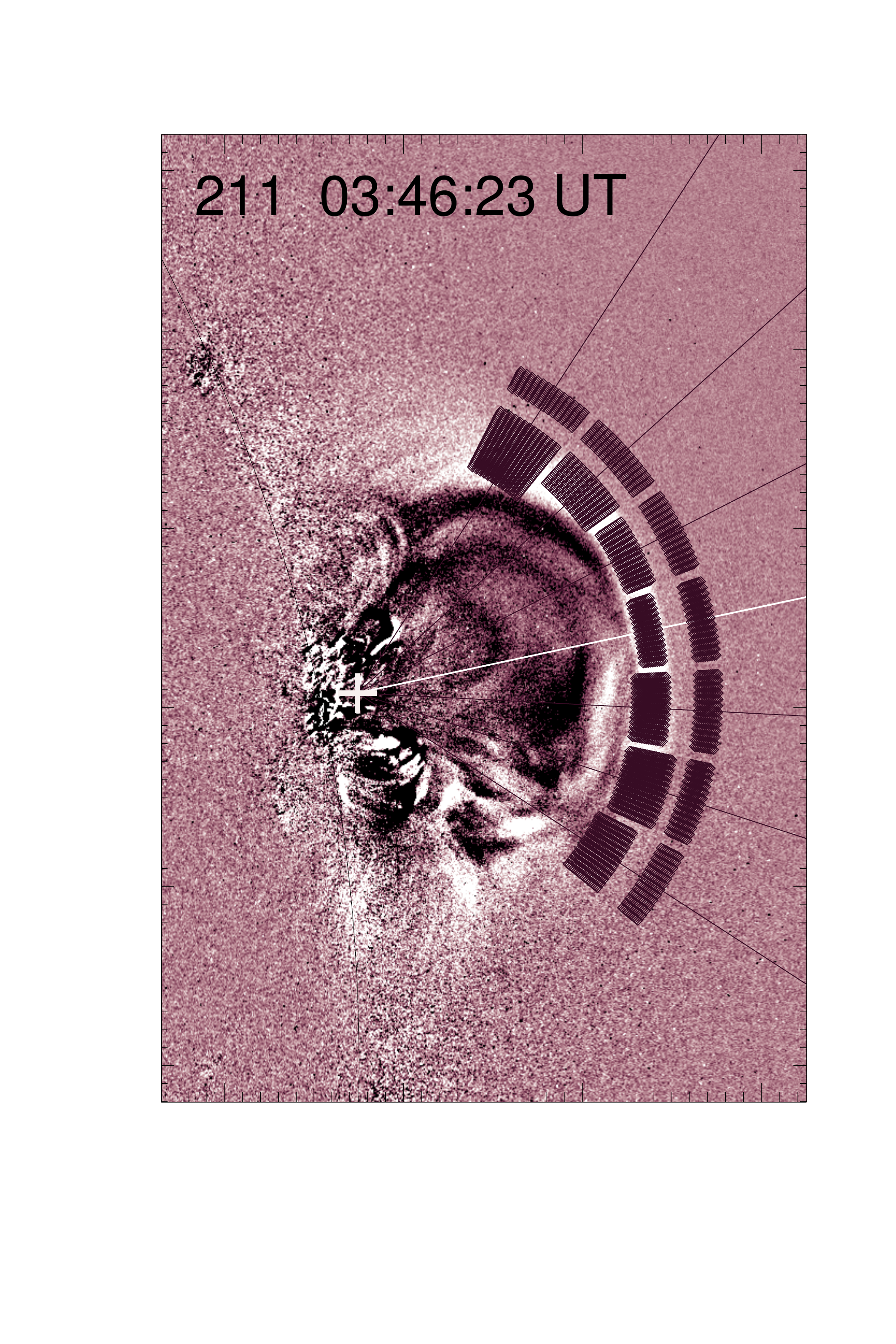}
\end{minipage}
\end{minipage}

\begin{minipage}[b]{\textwidth}
\centering
\begin{minipage}[b]{0.25\textwidth}
\includegraphics[width = 4.1cm,trim = 0 120 0 80,clip = true]{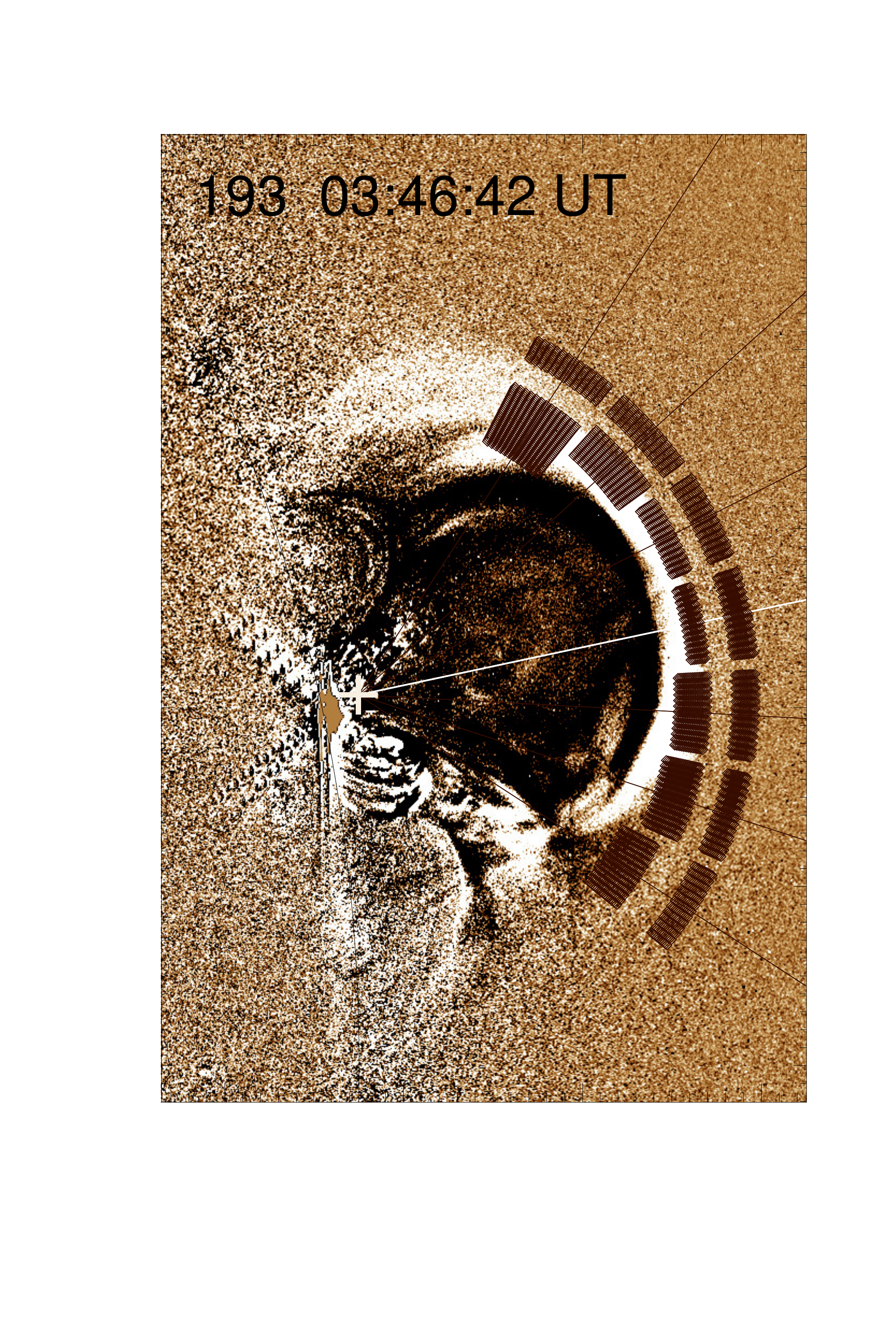}
\end{minipage}
\begin{minipage}[b]{0.25\textwidth}
\includegraphics[width = 4.1cm,trim = 0 120 0 80,clip = true]{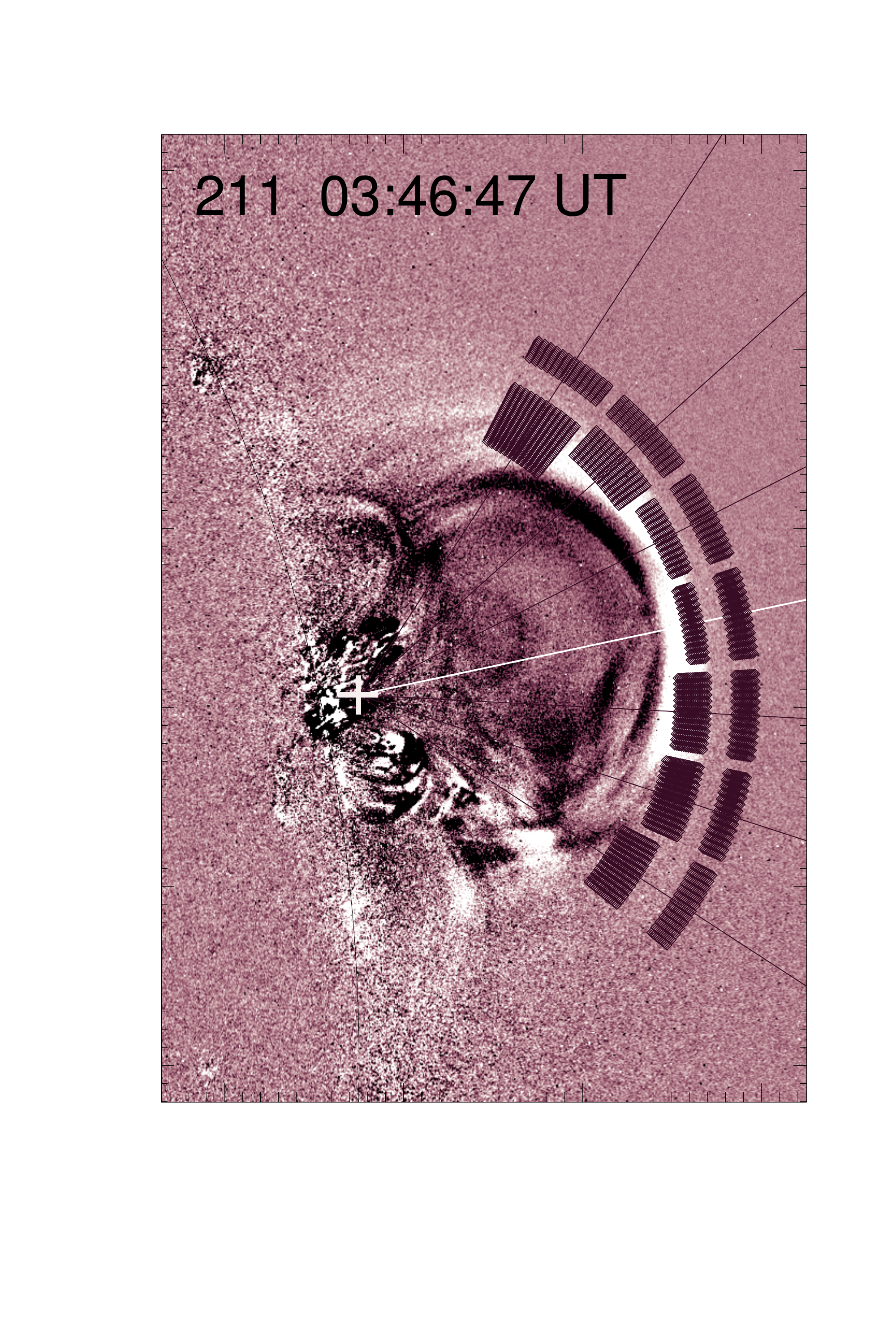}
\end{minipage}
\end{minipage}

\begin{minipage}[b]{\textwidth}
\centering
\begin{minipage}[b]{0.25\textwidth}
\includegraphics[width = 4.1cm,trim = 0 60 0 80,clip = true]{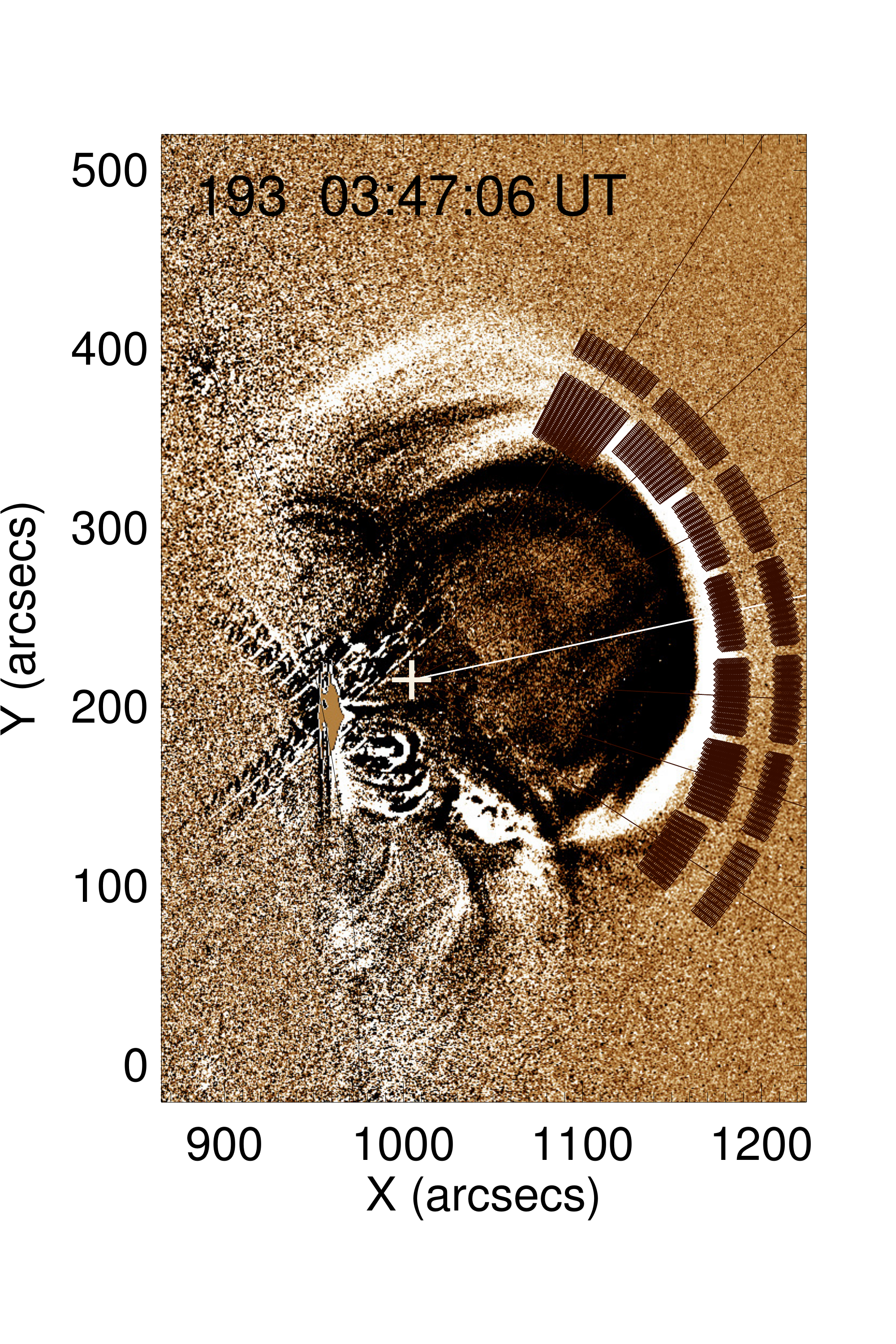}
\end{minipage}
\begin{minipage}[b]{0.25\textwidth}
\includegraphics[width = 4.1cm,trim = 0 60 0 80,clip = true]{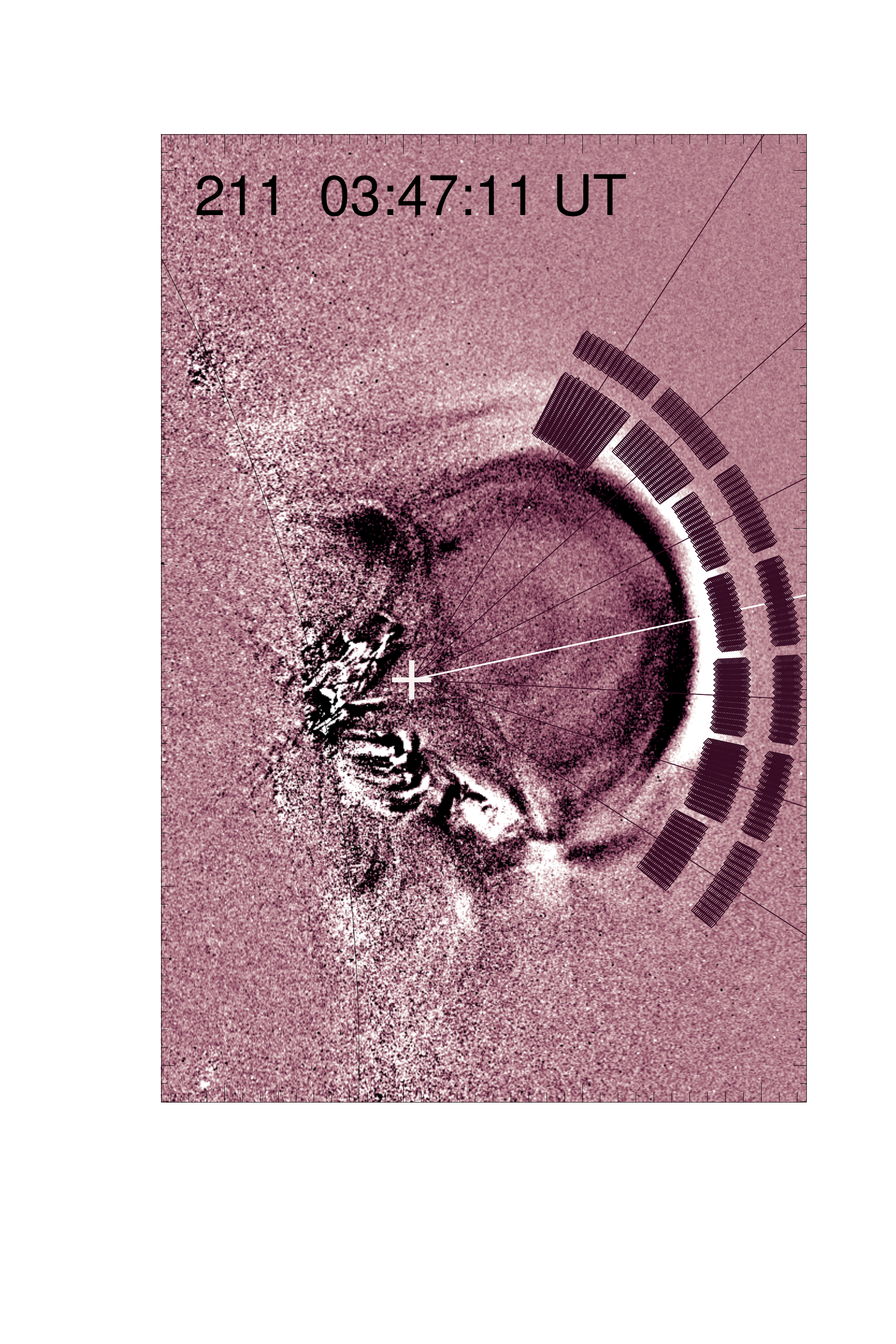}
\end{minipage}
\end{minipage}

\caption{\small{Running-difference images at 193 {\AA} (left) and 211 {\AA} (right) showing the evolution of the structure. The white pluses refer to the center of the fitted circle for the three moments. The line connecting the center of the solar disk and the center of the fitted circle is regarded as the base line (0$^{\circ}$), which is shown as white line. The OL is divided into 7 parts along directions of -45$^{\circ}$, -30$^{\circ}$, -15$^{\circ}$, 0$^{\circ}$, 15$^{\circ}$, 30$^{\circ}$, and 45$^{\circ}$ by the black solid lines. The inner shaded regions refer to disturbed regions of the OL, while the outer shaded regions to undisturbed regions of the OL.}}

\label{4-Alldirection}

\end{figure}

\begin{figure}[htbp]
\begin{center}
\plotone{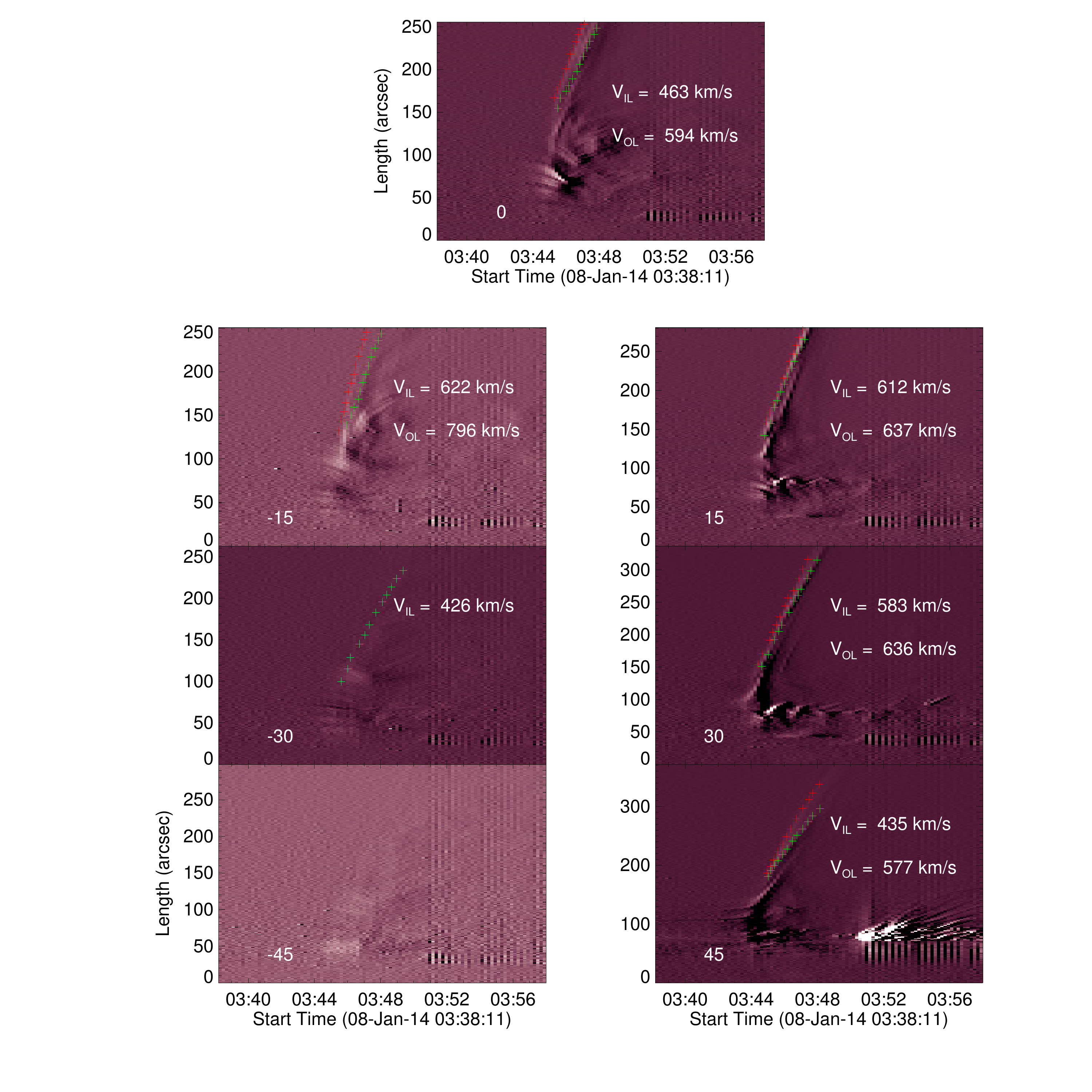}
\caption{\small{Time-slice plots in the 7 directions from the 211 {\AA} image. The red pluses refer to the upper edge of the bright structures, which corresponds to the outer layer, and the green pluses to the lower edge of the bright structures, which corresponds to the inner layer.}}
\label{4-V-211}
\end{center}
\end{figure}

\begin{figure}
  \center
   \includegraphics[width=12cm]{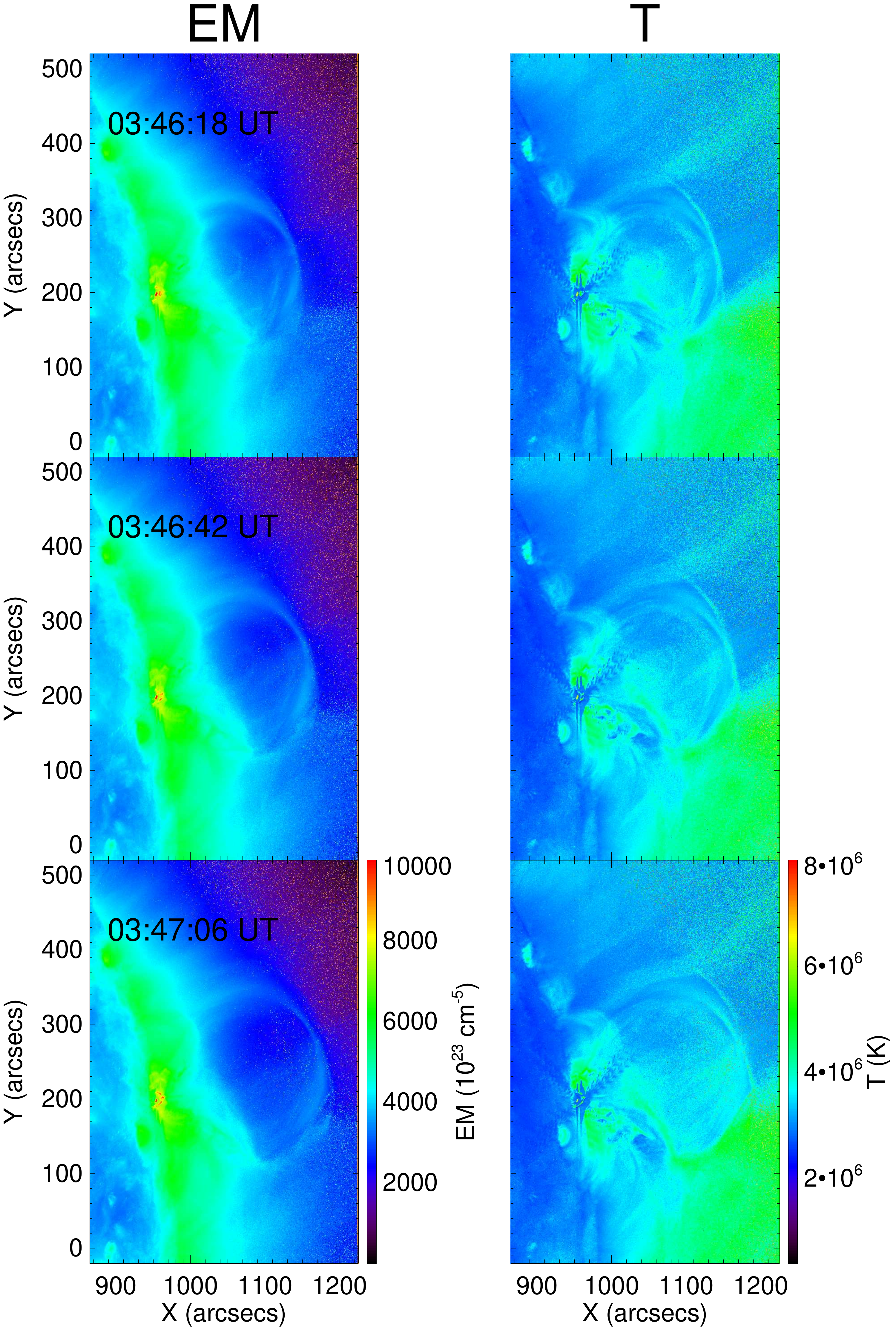}
   \caption{EM (left) and temperature (right) maps for the three moments in Fig. \ref{4-Alldirection}.}
   \label{4-denT}
\end{figure}

The DEM method can be applied to multi-wavelength data in order to derive the temperature and the EM values \citep{Weber2004,Cheng2012}. Here, we use the intensity images in 94 {\AA}, 131 {\AA}, 171 {\AA}, 193 {\AA}, 211 {\AA}, 335 {\AA} observed by {\it SDO}/AIA to derive the EM and the temperature distributions. Figure \ref{4-denT} shows the EM and the temperature maps at the three moments (see Figure~\ref{4-spec-LEAR-SSRT}). The double-layered structures also appear in the EM maps and temperature, similar to the structures in the EUV images as shown in Figure \ref{4-aia-overview}. The two layers are more distinct in the temperature maps. On the other hand, the noises in the temperature maps are larger than those in the EM maps.

\begin{figure}[htbp]
\begin{center}
\plotone{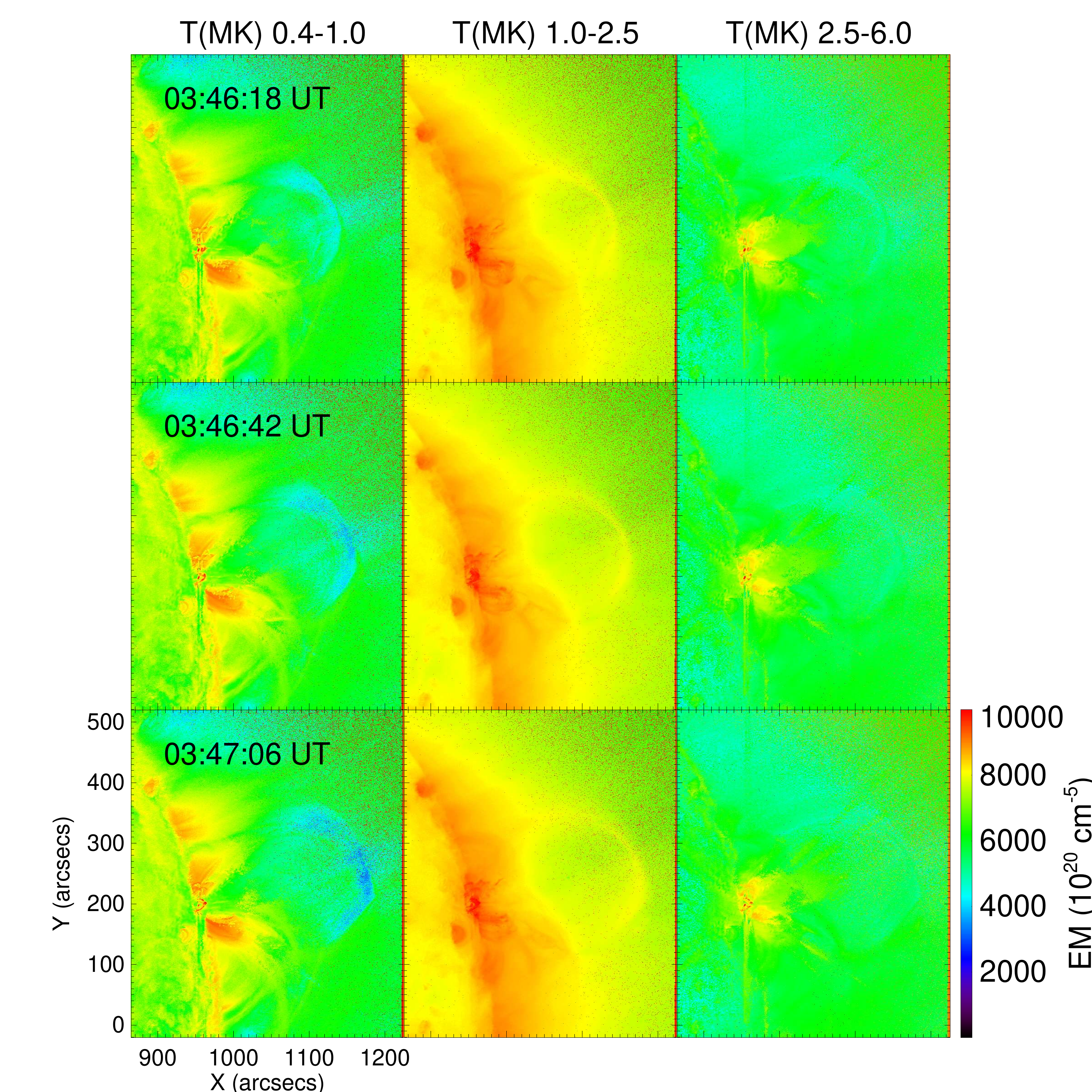}
\caption{\small{EM maps in different temperature ranges (0.4--1.0 MK, 1.0--2.5 MK, 2.5--6.0 MK) for the three moments in Fig. \ref{4-Alldirection}.}}
\label{4-EM-dt}
\end{center}
\end{figure}

In order to check the different properties of the OL in different temperature ranges during the OL propagation, we calculate the EM distributions in three temperature ranges 0.4--1.0 MK, 1.0--2.5 MK, 2.5--6.0 MK, as shown in Figure~\ref{4-EM-dt}. In the low temperature range 0.4--1.0 MK, the EM is the minimum in the OL region, but in the medium temperature range, 1.0--2.5 MK, the EM increases significantly in the OL, though not as strong as in the flaring site. In the high temperature range 2.5--6.0 MK, the EM has no obvious change. Such a result indicates a slight heating of the plasma around the OL.

In order to quantitatively study the properties of the different regions around the OL, we divide the OL into 7 segments in different directions as mentioned in Section~\ref{sect:Data Analysis}. In each direction, the density-enhanced region is considered to be the area disturbed by the OL, whereas the background corona on the outer side is considered as the undisturbed area. That is to say, the EM-enhanced region corresponds to the disturbed region and the outer background corona to the undisturbed region of the OL.

\begin{figure}
  \center
   \includegraphics[width=12cm]{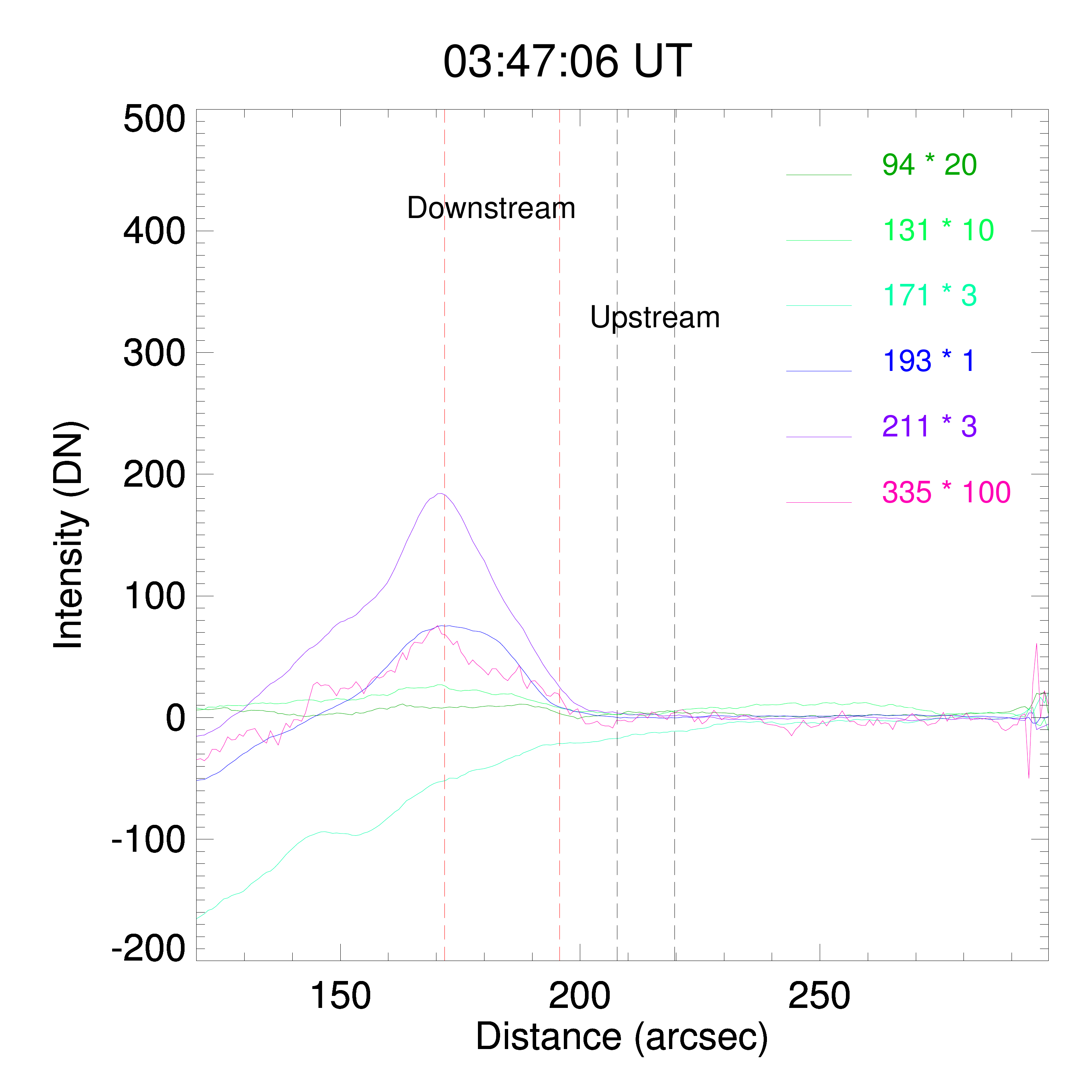}
   \caption{The intensity profiles in the 30$^{\circ}$ direction at 03:47:06 UT. The curves with different colors are base-difference intensities for different wavelengths, the base time is 03:36:06 UT. The region between the red dashed lines is regarded as the disturbed region, the region between black dashed lines is the undisturbed region.}
   \label{I034713T30}
\end{figure}

As shown in Figure~\ref{I034713T30}, the curves with different colors are base-difference intensities for different wavelengths in the 30$^{\circ}$ direction at 03:47:06 UT. One can see that the OL is identified most clearly in the 193 {\AA} and 211 {\AA} intensity profiles in Figure~\ref{I034713T30}. The position and thickness of the downstream and upstream regions are estimated from these intensity profiles. For example, the region between the maximum intensity and the normal level in the 193 {\AA} and 211 {\AA} intensity profiles is considered as the region that is disturbed by the OL, and the region where the intensity profiles is roughly flat is considered as the undisturbed region. To reduce the noises in the EM and the temperature maps, we set an angular width of $\pm$6$^{\circ}$ for each direction and make an average of the EM and temperature values azimuthally. The disturbed and undisturbed regions of the OL in each direction used in calculations are shown as the shaded areas in Figure~\ref{4-Alldirection}.

\begin{figure}
\center
\plotone{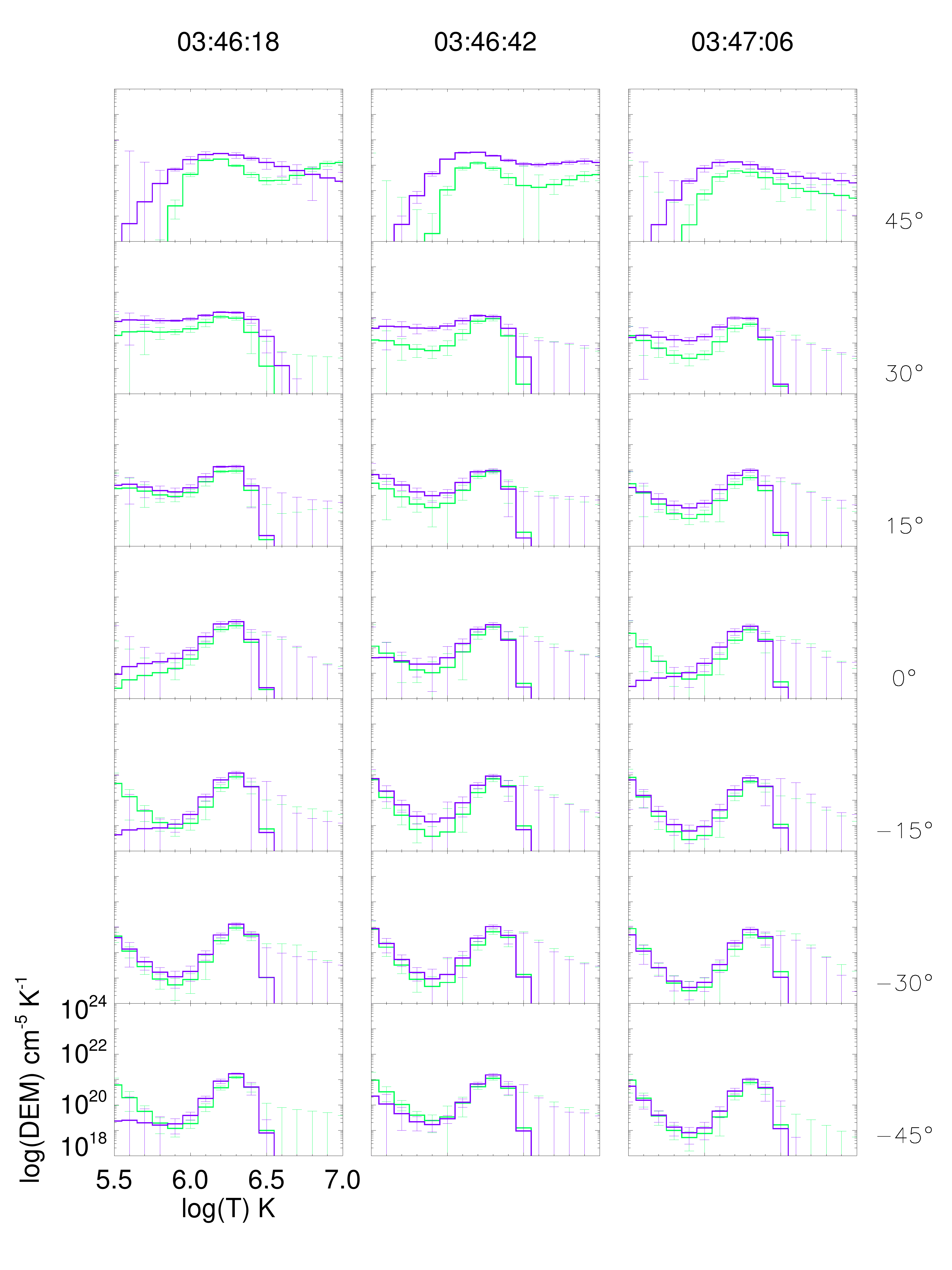}
\caption{\small{DEM results for the undisturbed (green) and disturbed (purple) regions of the OL in 7 directions at three moments of 03:46:18 UT, 03:46:42 UT and 03:47:06 UT.}}
\label{4-DEMcurve-all}
\end{figure}

Our calculations proceed as follows. First, we average the intensity of either the disturbed or the undisturbed region in each direction to a pixel.
We then apply the DEM method to the intensity values at different wavelengths to derive the EM and the temperature values in the disturbed and undisturbed regions of the OL. The DEM results are shown in Figure~\ref{4-DEMcurve-all}. One can notice that the DEM results of the disturbed and undisturbed regions of the OL are obviously different in the directions of 45$^{\circ}$. It indicates the quite different properties of the disturbed and undisturbed regions in this direction.

The electron density is usually calculated from the EM as \citep{Cheng2012}:
\begin{equation}
\label{equation3}
    n_{e} = \sqrt{\frac{EM}{l}},\\
\end{equation}
where $l$ is the geometrical depth along the line of sight (LOS). For optically thin emission, it is difficult to determine the LOS depth that contributes to the EM. Theoretically, the EM is an integration along the whole ray path. However, since the contribution per unit length drops significantly when moving away from the Sun,  the effective depth that contributes most of the EM is limited.
\citet{Zucca2014a} gave a formula to estimate the effective LOS depth:
\begin{equation}
\label{equation4}
    l \sim \sqrt{H \pi r},  \\
\end{equation}
where $H$ is the scale height, and $r$ is heliocentric distance. Note that this formula relies on the assumption that the EM per unit length is uniform. However, the EM per unit length is actually not uniform, especially for the situation when particular structures (e.g., pileup region, shock front, corona loops) run across the LOS.

\begin{figure}
\center
\plotone{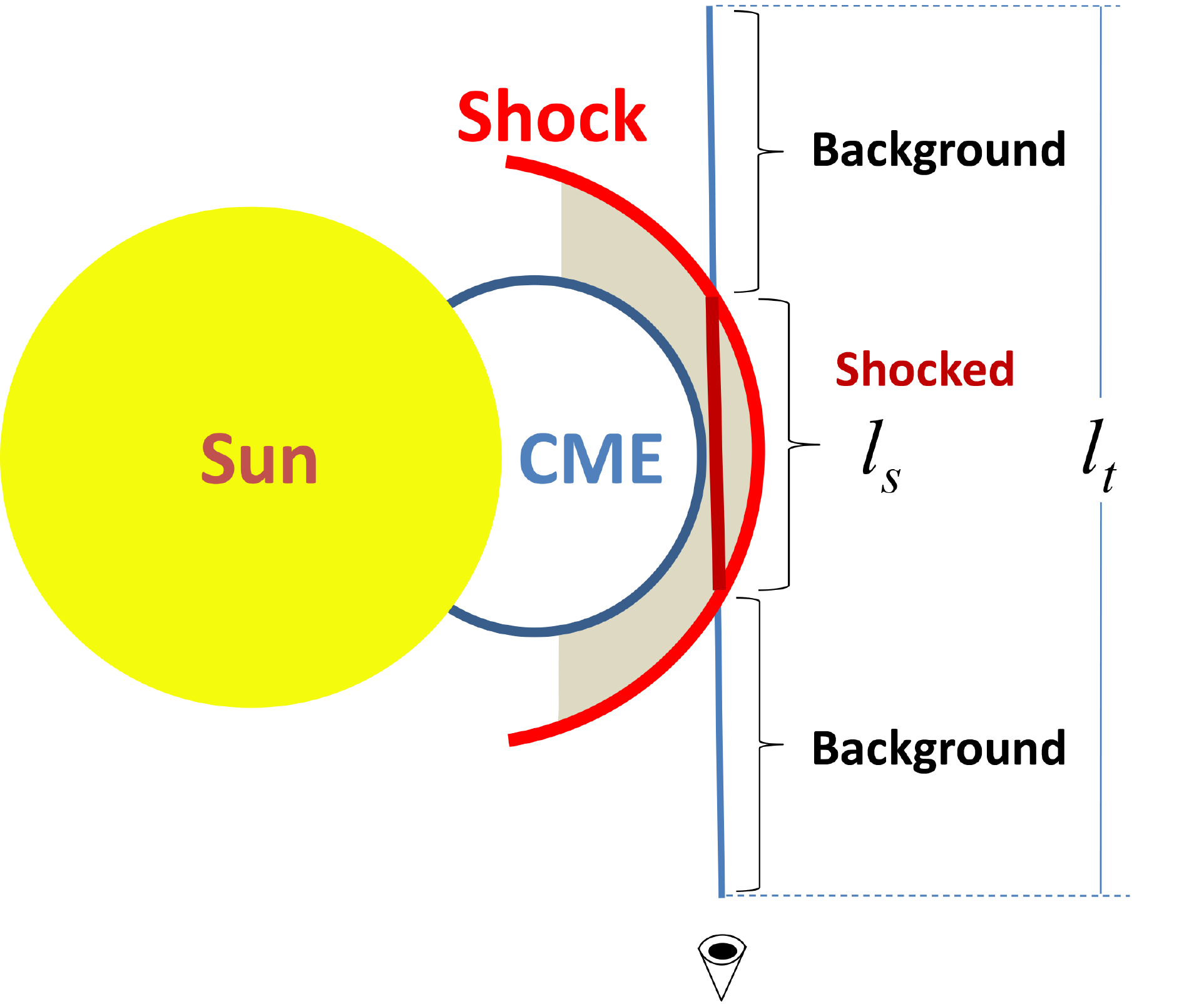}
\caption{\small{Schematic picture showing the effective depth contributing to the EM along LOS. When the shock front is just across the LOS, the total effective depth ($l_{t}$) consists of two parts, a shocked region ($l_{s}$) and an undisturbed region. The shocked region is considered as the sheath region between CME LE and shock front, which is indicated by the shaded region.}}
\label{4-line-of-sight}
\end{figure}

Here, we propose a revised method to estimate the effective LOS depth. Take a shock as an example,
when the shock front runs across the LOS, as shown in Figure~\ref{4-line-of-sight}, the effective depth contributing to the EM can be divided into two parts, one of which is the undisturbed region (background), the other is the region disturbed by the shock. Because of the different temperatures and the plasma densities in these two regions, their contributions to the EM per unit length are also different. Here, the total EM is denoted by $EM$, and the EM per unit length is denoted by $em$. Because the heliocentric distances of the downstream and upstream regions are similar, we assume that the total effective depth, $l_{t}$, is the same for both regions.

For the upstream region, the EM is expressed as
\begin{equation}
\label{equation5}
    EM_{0} = em_{u} l_{t}, \\
\end{equation}
where the subscript $u$ denotes the upstream region.

For the downstream region, at $t1$, when the shock runs across the LOS, the total EM consists of two parts, one from the region disturbed by the shock ($l_{s}$) and the other from the background ($l_{t} - l_{s}$):
\begin{equation}
\label{equation6}
    EM_{1}^{t1} = em_{b}^{t1} (l_{t} - l_{s}) + em_{s} l_{s} = em_{b}^{t1} (l_{t} - l_{s}) + EM_{s}, \\
\end{equation}
where $em_{b}$ is the EM per unit length of the background (as shown in Figure~\ref{4-line-of-sight}) along the LOS, $em_{s}$ is that of the region disturbed by the shock (shocked region). While at $t0$, the time before the shock runs across the LOS, the total EM is:
\begin{equation}
\label{equation7}
    EM_{1}^{t0} = em_{b}^{t0} l_{t}. \\
\end{equation}

Here, $em_{b}$ does not change before and after the shock passes across the LOS; thus, $em_{b}^{t0}$ is the same as $em_{b}^{t1}$. Combining the Equations (\ref{equation3}), (\ref{equation5}), and (\ref{equation6}), we get the compression ratio of the shock as
\begin{equation}
\label{equation8}
    X = \frac{n_{2}}{n_{1}} = \sqrt{\frac{EM_{s}/l_{s}}{EM_{0}/l_{t}}} = \sqrt{\frac{(l_{t}/l_{s}) \Delta EM + EM_{1}^{t0}}{EM_{0}}}, \\
\end{equation}
where \textbf{$\Delta EM$ \textbf{$(EM_{1}^{t1} - EM_{1}^{t0})$}} is the change of the total EM before and after the shock passing. The quantities $EM_{0}$, $EM_{1}$ and $\Delta EM$ can be obtained from the DEM method, and $l_{t}$ from Equation (\ref{equation4}).
We further adopt the depth of the shock $l_{s}$ as the diameter of the fitted circle \citep[also see,][]{Gopalswamy2013}. Thus, the quantity $X$ can be calculated from Equation (\ref{equation8}). As mentioned in Section 3.2, we are also able to estimate value of $M_{A}$ from Equation (\ref{equation2}).

In Table~\ref{4-table-direction-time}, we list the parameters of the OL, such as the emission measure, temperature ratio, compression ratio $X$, and Alfv\'{e}n Mach number $M_{A}$ in each direction for the three moments. From Table \ref{4-table-direction-time}, we find that the values of $X$ and $M_{A}$ of the OL in the directions of 45$^{\circ}$ are larger than those in other directions. The value of $X$ in the direction of 45$^{\circ}$, which is to 1.86 $\pm$ 0.13, 2.26 $\pm$ 0.34, and 1.78 $\pm$ 0.30 for the three moments, respectively. These values are slightly larger than those calculated from the dynamic spectrum mentioned in Section 3.2. Overall, the values of $X$ for the directions of 45$^{\circ}$ are comparable between the two different methods, considering the uncertainties in the measurements. This implies that the source region of the type II radio burst lies in the part of the OL around the direction of 45$^\circ$, where the OL front sharpens to become a shock wave locally. However, the OL front remains a pileup compression front in other directions.

\section{Discussion}
\label{sect:Discussion}

When applying the DEM method to calculate the electron density, the difficulty is how to estimate the effective LOS depth $l$. \citet{Zucca2014a} proposed a method to estimate the effective LOS depth by assuming the EM per unit length along the LOS is the same. But actually, the EM is not uniform, especially in the situation where different structures exist along the LOS.
For example, since the plasma is compressed by the shock, the EM along the LOS is more concentrated at the shock front rather than in the background. Therefore, the method should be improved by taking into account the inhomogeneity of the EM distribution. Base on the method of \citet{Zucca2014a}, we propose a more reasonable method to estimate the LOS depth by considering that the EM per unit length of a shock is different from that of the background along the LOS. We then derive Equation~(\ref{equation8}) to calculate the compression ratio of the shock. Such a method for estimating the effective LOS depth can also be used for other structures in the corona, such as coronal loops, coronal streams, and so on.

Here, we use two methods to measure the compression ratio of the shock front independently. One is based on the DEM method with an improved estimate of the effective LOS depth, and the other is based on the band-splitting in the dynamic spectrum of the type II radio burst. We found that the compression ratio in the direction of 45$^\circ$ derived from the DEM method is similar to that derived from the band-splitting. Such a result suggests that the source region of the type II radio burst is in the section of the OL front along the direction of 45$^\circ$. We also found that the compression ratio and the Alfv\'{e}n Mach number in this section are obviously larger than those in other directions. As mentioned in Section~3.1, the OL is the fastest in the direction of -15$^\circ$. That is to say, the source region of the type II radio burst is not in the section of the disturbance which travels the fastest. By comparison, some of the previous studies suggested that type II radio bursts are likely to be generated at the nose of the shocks because the shock speeds there are usually larger than at the flanks \citep{Bemporad2011}. For this event, however, the type II radio burst is probably generated in the segment of the disturbance where the compression ratio and the Alfv\'{e}n Mach number are the largest but the disturbance speed is not. It is known that the Alfv\'{e}n Mach number is a proxy of the strength of the disturbance, and a larger Alfv\'{e}n Mach number implies a stronger disturbance. The disturbance can sharpen to become a shock when the Alfv\'{e}n Mach number is large enough. Quantitatively, the Alfv\'{e}n Mach number of a disturbance is $M_{A} = v/v_{A}$, where $v$ is the speed of the disturbance and $v_{A}$ is the Alfv\'{e}n speed of the background. It implies that for a given disturbance, the shock is most probably to form near the site with the lowest Alfv\'en speed. This is consistent with \citet{Gopalswamy2009}, who found that type II radio bursts are generated at a height where the Alfv\'{e}n speed of background is low. Note that our result does not necessarily contradict with the results by \citet{Bemporad2011}, considering the fact that in a background corona where the Alfv\'en speed does not vary significantly in space, the nose part of the CME-driven shock has the largest Mach number. From this event, we found that the disturbance speed or the Alfv\'{e}n speed of the background alone cannot serve as a sufficient condition to produce the type II radio burst, while a suitable combination of them, or more strictly, a larger Alfven Mach number (and a larger compression ratio) is more likely the condition for the generation of the burst.

For a perpendicular shock, the relationship between the pre-shock and post-shock temperatures is \citep{Ma2011}:
\begin{equation}
\label{equation9}
     \frac{T_{d}}{T_{u}} = \frac{1}{X}[1 + (1 - \frac{1}{X} - \frac{X^2 - 1}{2 M^2_{A}})\gamma M^2]. \\
\end{equation}
Taking $\gamma$ = 5/3, $T$ = 2.4 MK, and the hydrogen to helium abundance ratio H/He = 9, we get the sound speed is 160 km s$^{-1}$, and the Mach number $M$ = 3.6. Using the value of $X$ and $M_{A}$ obtained from the band-splitting, the temperature ratio ($T_{d}/T_{u}$) for the three moments are 1.08 $\pm$ 0.12, 1.09 $\pm$ 0.08, 1.07 $\pm$ 0.08, respectively. At the shocked region (45$^\circ$), we also derive the temperature ratio from the DEM method, which are 1.04 $\pm$ 0.06, 1.09 $\pm$ 0.03 and 0.97 $\pm$ 0.06 for the three moments, respectively. The results are a sightly lower than those derived from the band-splitting, but the difference is still within the error range. The temperature ratio is a slightly larger than 1 at the shocked region, being consistent with the results by \citet{Kozarev2011}. It is noted that occasionally such a ratio is slightly smaller than unity, which is not realistic both for a shock or a compression front. Such a smaller value mainly results from the uncertainty of the DEM method.


\section{Conclusions}
\label{sect:Conclusions}

In this paper, we studied a typical type II radio burst associated with a CME-driven shock that occurred on the west limb of the Sun. There is a pileup region ahead of an erupting flux rope, the pileup region is a double-layered structure and propagates outward as seen in 193 {\AA} and 211 {\AA} passbands. The OL is well fitted by a standard circle arc, and we divided the circle into 7 directions from -45$^\circ$ to 45$^\circ$ with a separation of 15$^\circ$ between each. We measure the OL speed in each direction and found that the OL is the fastest in the direction of -15$^\circ$. We used the DEM method to derive the temperature, emission measure, temperature radio, compression ratio, and Alfven Mach number for each direction and each time. The main results are summarized as follows:

1. We proposed a new method to estimate the effective LOS depth by considering that the EM per unit length in the disturbed region along the LOS path is different from that in the undisturbed region, and this method can also be used for other structures in the corona, such as coronal loops, coronal streams, and so on. A more reasonable compression ratio is then derived using this method.

2. Through comparing the compression ratio derived by the DEM method with that derived by the band-splitting in the radio dynamic spectrum of the type II radio burst, we concluded that the source region of the type II radio burst is in the part of the OL at the direction of 45$^\circ$. It indicates that the OL front sharpens to become a shock wave at the direction of 45$^\circ$.

3. The type II radio burst is probably generated at the part of the disturbance where $M_{A}$ is the largest but not the part where the disturbance speed is the fastest. Thus, the disturbance speed alone cannot serve as a sufficient factor in the generation of a shock. Instead, a more appropriate factor is the Alfv\'{e}n Mach number, i.e., the generation of a shock requires a larger $M_{A}$ of the disturbance.

\acknowledgements

{\it SDO} is a mission of NASA's Living With a Star Program, and SOHO is a mission of international cooperation between ESA and NASA. This work was supported by the NSFC (grants 11303016, 11373023, 11333009, 11533005, 11203014, and 11025314), NKBRSF (grants 2011CB811402 and 2014CB744203) and Surface Fund of Jiangsu (grant BK20161618).





\begin{sidewaystable}
\centering
\caption{Temperature ($T$), $EM$, temperature radio ($T_{d}/T_{u}$), compression ratio ($X$) and Alfv\'{e}n Mach ($M_{A}$) for the upstream and downstream of the shock.}
\begin{tabular}{lccccccccc}
\\
\hline
 Time (UT) & Parameters &  &  &  & Directions &  &  &  \\
\hline
   &  & -45$^{\circ}$ & -30$^{\circ}$ & -15$^{\circ}$ & 0$^{\circ}$ & 15$^{\circ}$ & 30$^{\circ}$ & 45$^{\circ}$ \\
\hline
03:46:18    & T$_{d}$ (MK) & 1.97 $\pm$ 0.47 & 2.05 $\pm$ 0.60 & 1.96 $\pm$ 0.28 & 1.84 $\pm$ 0.34 & 1.77 $\pm$ 43 & 1.80 $\pm$ 0.29 & 2.40 $\pm$ 0.12 \\
            & T$_{u}$ (MK) & 2.05 $\pm$ 0.10 & 2.10 $\pm$ 0.09 & 2.05 $\pm$ 0.09 & 1.88 $\pm$ 0 & 1.77 $\pm$ 0.11 & 1.71 $\pm$ 0.07 & 2.30 $\pm$ 0.07\\
            & T$_{d}$/T$_{u}$ & 0.96 $\pm$ 0.24 & 0.97 $\pm$ 0.29 & 0.95 $\pm$ 0.14 & 0.97 $\pm$ 0.18 & 0.99 $\pm$ 0.25 & 1.05 $\pm$ 0.17 & 1.04 $\pm$ 0.06 \\
            & EM$^{t0}_{d}$ (10$^{26}$ cm$^{-5}$) & 14.63 $\pm$ 3.00 & 11.17 $\pm$ 6.06 & 10.18 $\pm$ 6.07 & 10.37 $\pm$ 4.00 & 12.03 $\pm$ 2.78 & 24.43 $\pm$ 2.48 & 47.23 $\pm$ 4.11 \\
            & EM$^{t1}_{d}$ (10$^{26}$ cm$^{-5}$) & 14.67 $\pm$ 2.20 & 11.18 $\pm$ 2.51 & 10.16 $\pm$ 5.64 & 10.39 $\pm$ 5.15 & 14.99 $\pm$ 3.69 & 26.78 $\pm$ 3.54 & 65.79 $\pm$ 9.52 \\
            & EM$_{u}$ (10$^{26}$ cm$^{-5}$) & 10.64 $\pm$ 2.49 & 8.09 $\pm$ 2.64 & 7.18 $\pm$ 2.47 & 6.83 $\pm$ 3.57 & 9.73 $\pm$ 2.25 & 13.22 $\pm$ 3.81 & 23.80 $\pm$ 1.86 \\
            & X & 1.17 $\pm$ 0.19 & 1.17 $\pm$ 0.37 & 1.19 $\pm$ 0.51 & 1.23 $\pm$ 0.49 & 1.35 $\pm$ 0.23 & 1.48 $\pm$ 0.23 & 1.86 $\pm$ 0.13 \\
            & M$_{A}$ & 1.13 $\pm$ 0.14 & 1.13 $\pm$ 0.28 & 1.14 $\pm$ 0.38 & 1.18 $\pm$ 0.37 & 1.27 $\pm$ 0.17 & 1.37 $\pm$ 0.18 & 1.73 $\pm$ 0.12 \\
\hline
03:46:42    & T$_{d}$ (MK) & 2.01 $\pm$ 0.06 & 2.07 $\pm$ 0.01 & 2.02 $\pm$ 0.01 & 1.90 $\pm$ 0.02 & 1.79 $\pm$ 0.13 & 1.73 $\pm$ 0.12 & 2.48 $\pm$ 0.04 \\
            & T$_{u}$ (MK) & 2.02 $\pm$ 0.08 & 2.14 $\pm$ 0.05 & 2.12 $\pm$ 0.03 & 1.99 $\pm$ 0.04 & 1.90 $\pm$ 0.10 & 1.84 $\pm$ 0.09 & 2.26 $\pm$ 0.06 \\
            & T$_{d}$/T$_{u}$ & 0.99 $\pm$ 0.05 & 0.97 $\pm$ 0.02 & 0.95 $\pm$ 0.01 & 0.95 $\pm$ 0.02 & 0.93 $\pm$ 0.09 & 0.93 $\pm$ 0.09 & 1.09 $\pm$ 0.03 \\
            & EM$^{t0}_{d}$ (10$^{26}$ cm$^{-5}$) & 13.07 $\pm$ 4.19 & 9.19 $\pm$ 4.55 & 7.79 $\pm$ 3.92 & 7.51 $\pm$ 3.57 & 9.98 $\pm$ 2.78 & 15.87 $\pm$ 3.37 & 57.32 $\pm$ 5.07 \\
            & EM$^{t1}_{d}$ (10$^{26}$ cm$^{-5}$) & 13.06 $\pm$ 5.07 & 9.24 $\pm$ 4.58 & 7.78 $\pm$ 3.91 & 7.46 $\pm$ 3.72 & 10.05 $\pm$ 3.18 & 18.44 $\pm$ 2.57 & 68.39 $\pm$ 3.37 \\
            & EM$_{u}$ (10$^{26}$ cm$^{-5}$) & 10.24 $\pm$ 3.76 & 6.24 $\pm$ 4.29 & 6.17 $\pm$ 4.82 & 5.77 $\pm$ 4.37 & 7.96 $\pm$ 3.09 & 8.92 $\pm$ 3.63 & 15.00 $\pm$ 4.51 \\
            & X & 1.13 $\pm$ 0.33 & 1.21 $\pm$ 0.56 & 1.12 $\pm$ 0.56 & 1.14 $\pm$ 0.55 & 1.12 $\pm$ 0.30 & 1.51 $\pm$ 0.33 & 2.26 $\pm$ 0.34 \\
            & M$_{A}$ & 1.09 $\pm$ 0.24 & 1.16 $\pm$ 0.42 & 1.09 $\pm$ 0.41 & 1.10 $\pm$ 0.41 & 1.09 $\pm$ 0.22 & 1.40 $\pm$ 0.26 & 2.17 $\pm$ 0.42 \\
\hline
03:47:06    & T$_{d}$ (MK) & 2.08 $\pm$ 0.34 & 2.12 $\pm$ 0.38 & 2.08 $\pm$ 0.32 & 1.92 $\pm$ 0.42 & 1.90 $\pm$ 0.29 & 1.75 $\pm$ 0.31 & 2.36 $\pm$ 0.13 \\
            & T$_{u}$ (MK) & 2.14 $\pm$ 0.12 & 2.18 $\pm$ 0.10 & 2.15 $\pm$ 0.17 & 2.05 $\pm$ 0.15 & 1.96 $\pm$ 0.12 & 1.88 $\pm$ 0.13 & 2.43 $\pm$ 0.08 \\
            & T$_{d}$/T$_{u}$ & 0.97 $\pm$ 0.17 & 0.96 $\pm$ 0.18 & 0.96 $\pm$ 0.17 & 0.94 $\pm$ 0.23 & 0.97 $\pm$ 0.17 & 0.92 $\pm$ 0.19 & 0.97 $\pm$ 0.06 \\
            & EM$^{t0}_{d}$ (10$^{26}$ cm$^{-5}$) & 8.99 $\pm$ 1.85 & 7.16 $\pm$ 5.36 & 6.56 $\pm$ 2.32 & 6.24 $\pm$ 3.67 & 7.00 $\pm$ 2.06 & 10.42 $\pm$ 1.85 & 27.96 $\pm$ 4.54 \\
            & EM$^{t1}_{d}$ (10$^{26}$ cm$^{-5}$) & 9.10 $\pm$ 3.65 & 7.18 $\pm$ 3.77 & 6.63 $\pm$ 5.02 & 6.31 $\pm$ 3.48 & 8.84 $\pm$ 5.23 & 10.70 $\pm$ 3.26 & 30.71 $\pm$ 3.52 \\
            & EM$_{u}$ (10$^{26}$ cm$^{-5}$) & 7.14 $\pm$ 2.39 & 5.12 $\pm$ 4.00 & 5.03 $\pm$ 1.73 & 4.44 $\pm$ 2.52 & 4.44 $\pm$ 3.22 & 5.17 $\pm$ 3.05 & 10.38 $\pm$ 3.31 \\
            & X & 1.13 $\pm$ 0.30 & 1.18 $\pm$ 0.66 & 1.15 $\pm$ 0.50 & 1.19 $\pm$ 0.55 & 1.52 $\pm$ 0.68 & 1.45 $\pm$ 0.49 & 1.78 $\pm$ 0.31 \\
            & M$_{A}$ & 1.10 $\pm$ 0.23 & 1.14 $\pm$ 0.49 & 1.11 $\pm$ 0.37 & 1.14 $\pm$ 0.41 & 1.41 $\pm$ 0.54 & 1.35 $\pm$ 0.38 & 1.65 $\pm$ 0.28 \\
\hline
\end{tabular}
\label{4-table-direction-time}
\end{sidewaystable}

\end{document}